\documentclass[12pt]{article}
\usepackage{graphicx} 
\usepackage{amsmath,amssymb}
\usepackage{mathtools}
\usepackage{subcaption}
\usepackage[colorlinks=true,linkcolor=blue,citecolor=blue,urlcolor=blue]{hyperref}
\usepackage{lineno}
\usepackage{authblk}
\usepackage{cite}
\usepackage[italicdiff]{physics}
\providecommand{\keywords}[1]{\noindent\textbf{Keywords:} #1}

\title{Probing Cosmic Strings via Black Hole Quasinormal Modes in Gravitational Wave Astronomy}
\author[1]{Ishan Swamy}
\author[1]{Deobrat Singh \thanks{Corresponding author: deobrat.singh@mitwpu.edu.in}}

\affil[1]{Dr. Vishwanath Karad MIT World Peace University, Pune, India}
\date{}

\begin{document}

\maketitle

\abstract{Black holes, the simplest solution to Einstein's field equations, do not emit light, making their observations a major challenge for researchers. However, discovery of binary black holes (BBHs) in 2015 by LIGO has transformed the study of compact objects, with over 300 BBHs recorded, providing a new avenue for probing new physics. GWs remain a prominent and precise method of observing not only BBHs, but also dark matter and cosmic strings. Cosmic strings --- hypothetical one dimensional topological defects formed in the early universe, are yet to be observed, with multiple detection methods such as particle radiation, gravitational waves and lensing being proposed. Here we present a novel framework to search for cosmic strings by modeling them as perturbations within non-rotating black hole spacetime, focusing on their imprint on the spectrum of quasinormal modes (QNMs). Our numerical simulations identify a lower limit on perturbation strength, $\lambda \sim 10^{-10}$ for uncharged string and $\lambda \sim 10^{-7}$ for charged string, below which cosmic string effects become unobservable in QNM signals. By analyzing eigenvalue splitting and centers, we show that cosmic string properties impart distinct and detectable features to GW signals. Our results establish QNM analysis as a powerful, alternative observational strategy for constraining or detecting cosmic strings, and offer an inverse approach to estimate string energy or charge if a signal is detected. With upgrades in LIGO technologies and advanced multimessenger astronomy under development, these findings highlight new potential for detecting cosmic strings.}


\section{Introduction}
\label{intro}
Black Holes arise as the simplest solution to Einstein's equation of general relativity \cite{Einstein16} \cite{Schwarzs16}, yet these compact objects continue to puzzle researchers till date. With black holes proposed to emit no light or radiation, their presence was first inferred indirectly through gravitational influences on nearby matter—an approach exemplified by the 1964 identification of the black hole candidate in the X-ray binary Cygnus X-1, nearly five decades after the theoretical prediction. However, this indirect method was limited to identifying black holes in binary systems. \\

The breakthrough came with the direct detection of gravitational waves (GWs)—ripples in the fabric of spacetime generated by accelerating massive bodies such as merging binary black holes (BBHs) or neutron stars. A century after its prediction, the first direct evidence of black holes was recorded in 2015 by LIGO \cite{abbott16}. Since then, over 300 BBH mergers have been observed, revealing not only their existence but also fundamental black hole properties including masses and spins, thereby offering a powerful tool for testing gravitational theory and probing new physics. An example of this is a study \cite{Abac25} by the LIGO-Virgo-KAGRA collaboration on the gravitational-wave signal GW250114 which has been used to test existing theories of black hole thermodynamics, with results strongly in support of the Hawking area theorem \cite{Hawking71}. \\

GWs observations, however, aren't restricted to BBHs, with recent studies suggesting analysing and detecting objects such as dark matter \cite{Nakama20, Freese23,Bertone24, Maurya24, Zhang24,Jana25} and cosmic strings \cite{Buch21, Chang22, Josu24, Sousa24, Chitose25, Wang25} ---  topological defects predicted to emit gravitational radiation as they evolve and interact\cite{kibble76, Vachaspati85, caldwell92, damour00, siemens07, Matsunami19, Baeza24}, While cosmic strings have yet to be conclusively observed, their gravitational wave signatures have attracted significant theoretical and observational interest, with ongoing efforts to constrain their properties using data from GW observatories \cite{Aasi14, Auclair20, Abbott21}. \\

Cosmic strings are also implicated in diverse cosmological phenomena including particle radiation \cite{Davis86, Vilenkin86, Vilenkin87, Saurabh20} and large-scale structure formation \cite{Turok84,  Silk1984, Vachaspati1991, Shlaer2012, Jiao23, Jiao2024}. Their interactions with black holes have proved to be a fascinating domain \cite{Aryal86, Larsen93, Larsen94, Larsen96, Larsen98, Frolov99, Dubath07, Amjad21, Kinoshita16, Xing21, ahmed24}, with the authors also having studied this interaction previously in X-ray binary systems \cite{Swamy25, singh25, Swamyprd}. \\

In this work, we bridge the fields of GW astronomy, cosmic string physics, and black hole perturbation theory by examining how charged and uncharged cosmic strings affect the QNM spectra of Schwarzschild black holes as a method to detect cosmic strings. QNMs are characteristic damped oscillations of black holes that manifest due to perturbations in black hole spacetime \cite{Vishveshwara70, Press71, Teukolsky73, Chandrasekhar75, kokkotas99, nollert93, Leaver85} which provide significant information about the remnant black hole's properties. We introduce two types of cosmic strings, charged and uncharged, as perturbations for the case of Schwarzschild black hole and attempt to understand how it effects the QNM signals. Although there have been analytical approaches to QNMs and perturbations \cite{Heidari23, Motohashi25}, cosmic strings are considered and numerically analysed as perturbations for the first time in this work. \\

The structure of this study is as follows: Section \ref{maths} describes the system and its corresponding equations. In Section \ref{analysis} we numerically analyse the system by estimating and comparing the eigenvalues for unperturbed and perturbed cases, including both charged and uncharged strings. We then proceed with calculating the eigenvalue centers and splitting for multiple perturbation strengths and illustrate how the presence of cosmic string affects QNMs, emphasizing its importance in GW observations. Finally, we present and explain our results in Section \ref{results}, highlighting how this work provides a novel framework for detecting cosmic strings using GWs. We use natural units with $G=c=\hbar=1$ throughout.

\section{Perturbed Schwarzschild Hamiltonian}
\label{maths}
We consider a Schwarzschild black hole with metric defined \cite{Schwarzs16} as 
 \begin{equation}
\mathrm{d}s^2 = -\left(1 - \frac{2M}{r}\right)\mathrm{d}t^2 + \left(1 - \frac{2M}{r}\right)^{-1}\mathrm{d}r^2 + r^2\mathrm{d}\theta^2 + r^2sin^2(\theta)\mathrm{d}\phi^2
\label{metric}
\end{equation}
where $t, r , \theta, \phi$ are Schwarzschild coordinates and M is the mass of the black hole.

Restricting to zero angular momentum photons confined to the equatorial plane, the classical Hamiltonian of the system reads \cite{Misner73}, \cite{Chandra83},
\begin{equation}
H
= \frac{1}{2}\!\left[
\left(1 - \frac{2M}{r}\right)p_r^2 
+ \frac{p_{\phi}^2}{r^2} 
- \frac{E_p^2}{\left(1 - \frac{2M}{r}\right)}
\right] = 0.
\label{hamiltonian}
\end{equation}
with constraint $H = 0$. The term $-E_p^2/\!\left[2(1 - 2M/r)\right]$ originates from the conserved energy $E_p$ of the test particle.

Now we attempt to convert this Hamiltonian to an operator format using 
\begin{equation}
p_r \rightarrow \hat{p_r} = -i\hbar\pdv{r}, \quad
p_\phi \rightarrow \hat{p_\phi} = -i\hbar\pdv{\phi}
\label{operator}
\end{equation}
which ultimately leads to a time-independent Schrödinger-like equation after an appropriate operator ordering, consistent with \cite{Regge57}, \cite{Schutz85},
\begin{equation}
\hat{H} = -\hbar^2\left(1-\frac{2M}{r}\right)\pdv[2]{r} - \frac{\hbar^2}{r^2}\pdv[2]{\phi}
\label{hamop}
\end{equation}

To remove the coordinate singularity at $r = 2M$ we introduce a tortoise coordinate $r_* = r+ 2M\ln|r-2M|$ such that 
\begin{equation}
\dv{r_*}{r} = \left(1-\frac{2M}{r}\right)^{-1}
\label{tort}
\end{equation}
mapping the radial coordinate to $-\infty<r_*<\infty$, conforming to Regge and Wheeler's approach \cite{Regge57}.

Thus, the Schrodinger-like equation for Schwarzschild black hole becomes
\begin{equation}
-\hbar^2\pdv[2]{\psi}{r_*} + V_{\rm eff}(r)\psi = \omega^2\psi
\label{schrodinger}
\end{equation}
where the effective potential corresponds to the Regge–Wheeler potential
\begin{equation}
V_{\rm eff}(r) = \left(1-\frac{2M}{r}\right) \frac{l(l+1)}{r^2}
\label{effpot}
\end{equation}
with $l$ the angular momentum quantum number, matching with \cite{Regge57, Misner73}.
Here, $\omega$ represents the frequency (or energy eigenvalue) of the wave mode obtained from the time-translation symmetry of the Schwarzschild background. In the geometric-optics limit, this quantity corresponds directly to the conserved particle energy $E_p$ that appears in the classical null-geodesic Hamiltonian, where $p_t = -E_p$. Thus, in the absence of perturbations, one may regard $\omega \simeq E_p$. However, it is important to distinguish this dynamical energy of the perturbation mode from the string energy parameter $E_s$ introduced later in \ref{ucspot} and \ref{ccspot}, which characterizes the intrinsic energy of the cosmic string (normalized by its tension) and acts as a property of the perturbing source rather than the black-hole mode itself.
 \\

Now we introduce a cosmic string in the equatorial plane as perturbation in the Schwarzschild spacetime with potential estimated \cite{Frolov99} as, 
\begin{equation}
V_{\rm UCS} = \frac{1}{2}r^2 - \frac{E_s^2}{2}\left(1-\frac{2M}{r}\right)^{-1} 
\label{ucspot}
\end{equation}
for an uncharged cosmic string, where $E_s$ is the total conserved energy of the string divided by $2 \pi \mu$ ($\mu$ being the string tension) and 
\begin{equation}
V_{\rm CCS} = \sqrt{r^2-2Mr}\left(1+\frac{N^2}{2r^2}\right) - \frac{E_s^2}{2}\left(1-\frac{2M}{r}\right)^{-1} 
\label{ccspot}
\end{equation}
for a charged cosmic string \cite{Larsen94} with an additional parameter $N$ corresponding to  the electromagnetic self-interaction energy accounting for electric charge contributions to the potential.

\section{Numerical Analysis}
\label{analysis}
In this section, detailed numerical simulations were performed to examine the impact of cosmic string perturbations—both uncharged and charged—on the quasinormal mode (QNM) spectrum of a Schwarzschild black hole using \ref{schrodinger}. The system studied consists of a black hole of 10 $M_\odot$ and string with energy term $E_s=3M$. In the range $r \in (2M,100M)$ its eigenvalues are computationally calculated for a fixed angular momentum number $l = 2$ and for 25 radial overtones $n$.  While here we focus on the fundamental modes with low $l =2$, higher $l$ modes are analyzed in Appendix \ref{appendix}.

\subsection{Uncharged Cosmic String Perturbations}

\begin{figure}[h!]
\begin{subfigure}{0.5\linewidth}
\centering
\includegraphics[scale = 0.45]{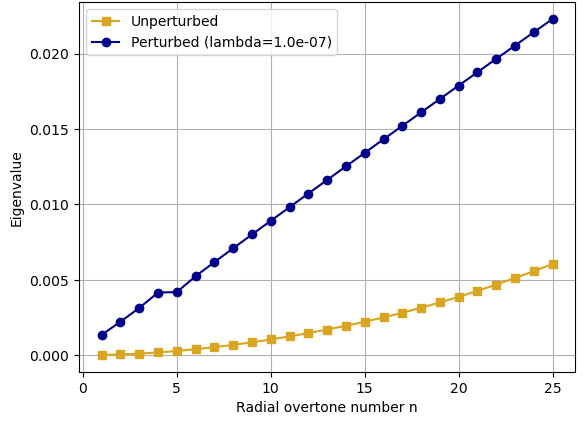}
\subcaption{$\lambda = 10^{-7}$}
\label{fig1a}
\end{subfigure}
\begin{subfigure}{0.5\linewidth}
\centering
\includegraphics[scale = 0.45]{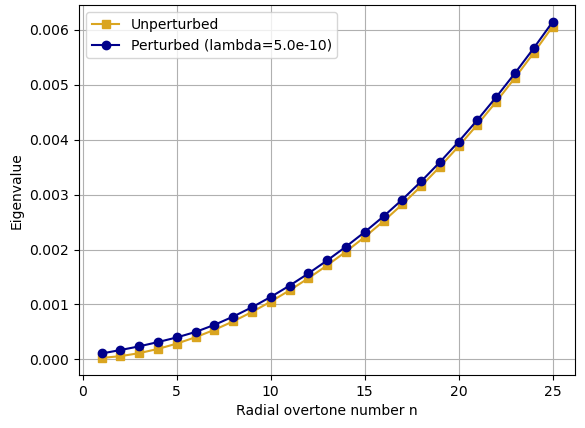}
\subcaption{$\lambda = 5 \times 10^{-10}$}
\label{fig1b}
\end{subfigure}
\caption{Perturbed vs Unperturbed eigenvalue comparison for uncharged cosmic string perturbation}
\label{fig1}
\end{figure}

We add a perturbation potential term corresponding to the uncharged cosmic string such that $V \rightarrow V + \lambda V_{\rm UCS}$ where $\lambda$ is a parameter controlling the strength of perturbation. Fig. \ref{fig1a} shows the plot for eigenvalues vs $n$ comparing the unperturbed (gold square point line) and perturbed (blue circle point line) case for a fixed value of $\lambda \sim 10^{-7}$. Further, Fig. \ref{fig1b} reveals that even an extremely small perturbation parameter, as low as $10^{-10}$, can produce pronounced shifts in the QNM eigenvalues. The comparison between perturbed and unperturbed spectra demonstrates that the cosmic string's presence is not only theoretically significant but also potentially observable, as such minute perturbations can become detectable with high-precision gravitational wave measurements. Increasing the perturbation strength consistently amplifies the shift in eigenvalues, enabling clear differentiation from the baseline black hole spectrum.

The analysis advances by treating two closely spaced QNM modes as a two-level system under perturbation, allowing calculation of eigenvalue splitting and centers \cite{Motohashi25}. The QNM frequencies that, after perturbation, get shifted $\omega_{n_i} \rightarrow \omega_\pm$ can be approximated as 

\begin{equation}
\omega_{\pm}^2 = \mathcal{E} \pm \sqrt{\mathcal{E}_d^2 + \Delta^2}
\label{eigenvalue}
\end{equation}
where $\mathcal{E}_{c,d} \coloneqq \frac{\mathcal{E}_{n_1} \pm \mathcal{E}_{n_2}}{2}, \quad
\mathcal{E}_{n_i} \coloneqq \omega_{n_i}^2 + \delta \omega_{n_i}^2$ for  i=1,2, and $\Delta \coloneqq ( \tilde{\psi}_{n_1} | \delta V | \psi_{n_2} )$
with $\psi_{n_i}$ normalized. \\

\begin{figure}[h!]
\begin{subfigure}{0.5\linewidth}
\centering
\includegraphics[width = \textwidth]{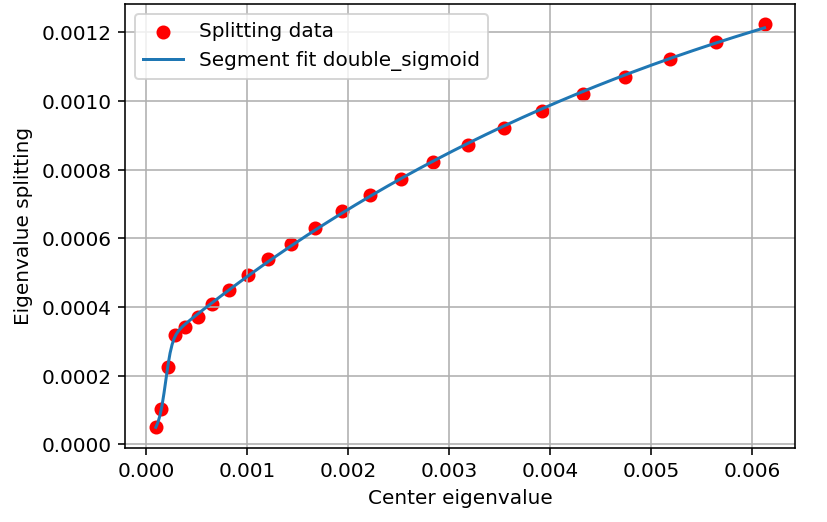}
\subcaption{$\lambda = 2.1 \times 10^{-10}$}
\label{fig2a}
\end{subfigure}
\begin{subfigure}{0.5\linewidth}
\centering
\includegraphics[width = \textwidth]{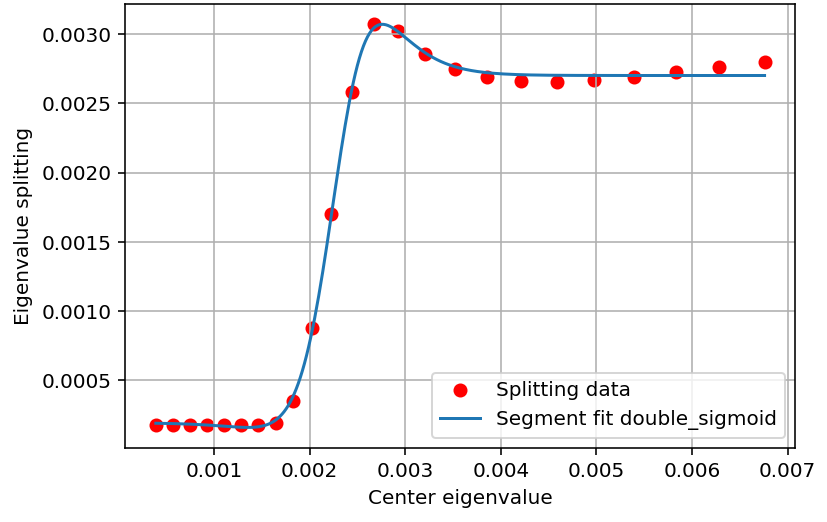}
\subcaption{$\lambda = 4.1 \times 10^{-9}$}
\label{fig2b}
\end{subfigure}
\begin{subfigure}{0.5\linewidth}
\centering
\includegraphics[width = \textwidth]{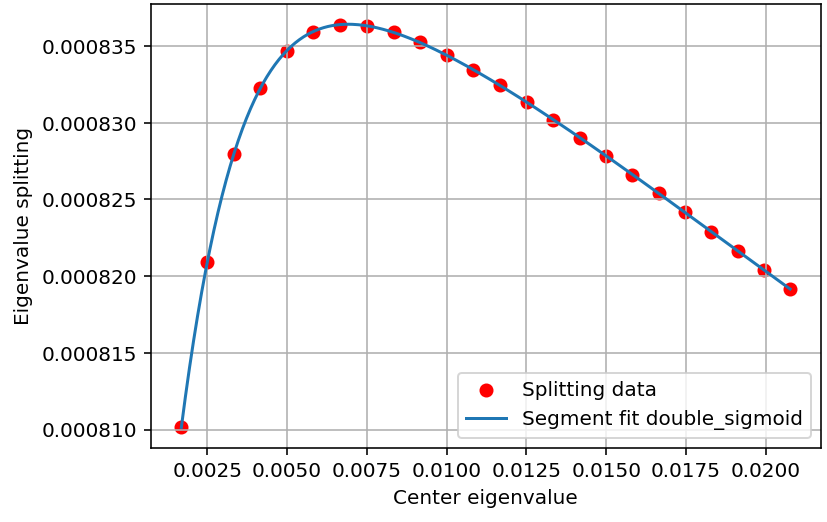}
\subcaption{$\lambda = 8.7 \times 10^{-8}$}
\label{fig2c}
\end{subfigure}
\begin{subfigure}{0.5\linewidth}
\centering
\includegraphics[width = \textwidth]{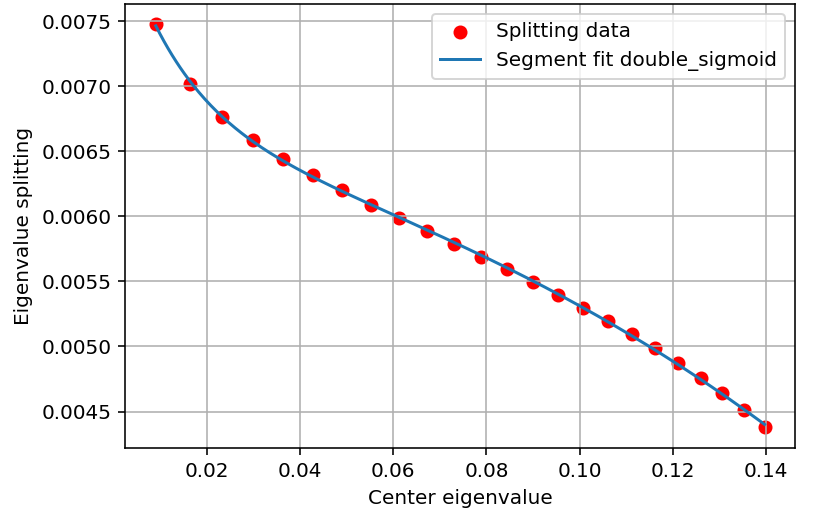}
\subcaption{$\lambda = 4.7 \times 10^{-6}$}
\label{fig2d}
\end{subfigure}
\begin{subfigure}{0.5\linewidth}
\centering
\includegraphics[width = \textwidth]{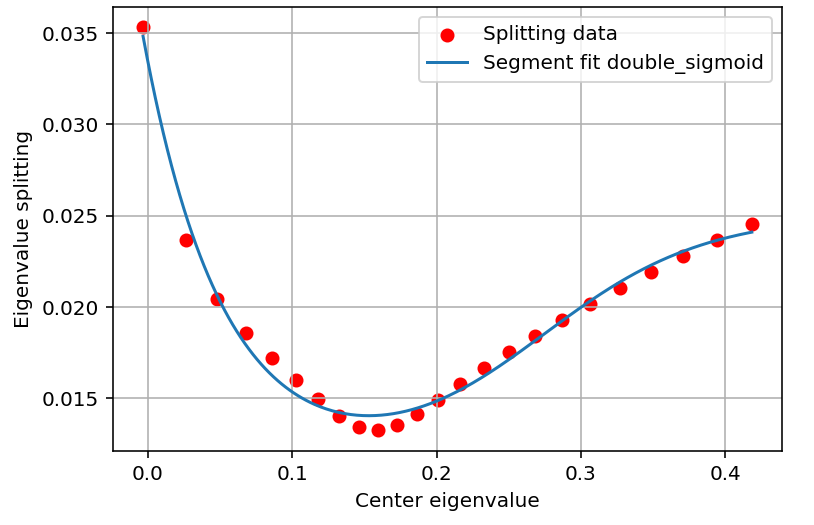}
\subcaption{$\lambda = 3.6 \times 10^{-5}$}
\label{fig2e}
\end{subfigure}
\begin{subfigure}{0.5\linewidth}
\centering
\includegraphics[width = \textwidth]{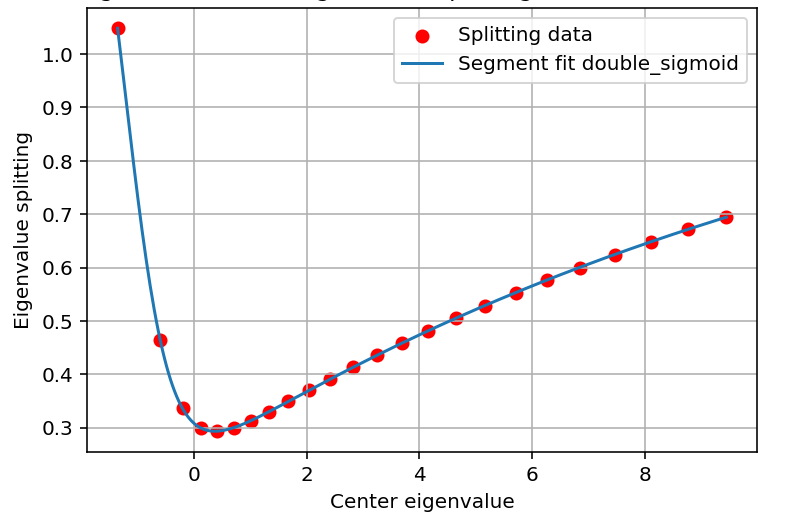}
\subcaption{$\lambda = 10^{-3}$}
\label{fig2f}
\end{subfigure}
\caption{Eigenvalue splitting vs center eigenvalue for the case of uncharged cosmic string perturbation for different lambda values.}
\label{fig2}
\end{figure}

The eigenvalue centers (x-axis) and eigenvalue splitting (y-axis) were plotted for 50 values of the perturbation strength $\lambda$  between $10^{-10}$ and $10^{-3}$ as shown in Fig. \ref{fig2}. Fitting this splitting-centers data to a double sigmoid function reveals systematic trends, yielding a coefficient of determination $R^2 > 0.95$ indicating an excellent fit. A significant feature of these plots is the dynamic pattern of the peak arising in the lower-left region of Fig \ref{fig2a} and evolving into distinctive peaks in Fig. \ref{fig2b} and Fig. \ref{fig2c}, occurring at different values of eigenvalue center and splitting. As $\lambda$ increases further, this peak transitions into a minimum, starting from the right of Fig. \ref{fig2d}, moving from the right side towards the left in Fig. \ref{fig2e} and Fig. \ref{fig2f}. A key observation is the clear differentiation of these plots for different $\lambda$ values, which is crucial for interpreting and constraining quasinormal mode properties in gravitational wave observations.

\subsection{Charged Cosmic String}

The system with charged cosmic string can be considered in the same way as the uncharged case by using \ref{ccspot} instead of \ref{ucspot} as the perturbation. For charged cosmic strings, however, the perturbation introduces an additional control parameter ($N$) linked to the string's electromagnetic energy.  The rest of the variables remain the same as in the uncharged string case. We consider three cases for $N$: $N^2 < N^2_{\rm crit}$; $N^2 = N^2_{\rm crit}$ and $N^2 > N^2_{\rm crit}$ where $N_{\rm crit}^2 = (13\sqrt{13} + 47)M^2$ corresponds to the minimum $N$ for which critical points (stable and unstable) outside the horizon exist \cite{Larsen94}.\\

\begin{figure}[ht]
\begin{subfigure}{0.5\linewidth}
\centering
\includegraphics[scale = 0.45]{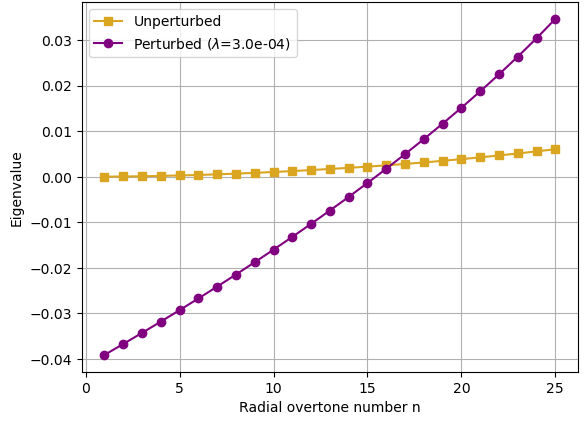}
\subcaption{$\lambda = 10^{-4}$}
\label{fig3a}
\end{subfigure}
\begin{subfigure}{0.5\linewidth}
\centering
\includegraphics[scale = 0.45]{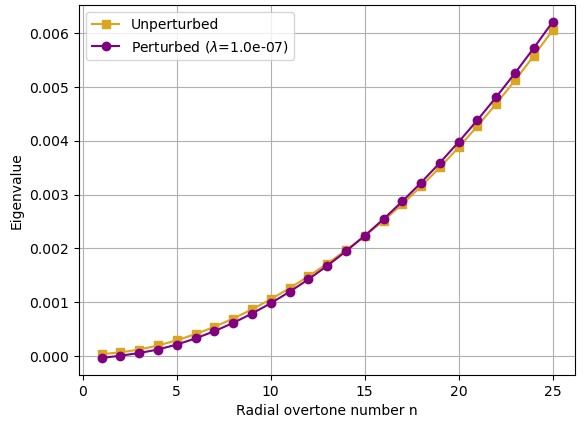}
\subcaption{$\lambda = 5 \times 10^{-7}$}
\label{fig3b}
\end{subfigure}
\caption{Perturbed vs Unperturbed eigenvalue comparison for charged cosmic string perturbation}
\label{fig3}
\end{figure}

We consider $N^2 = 0.5 N^2_{crit} = 48.44$ and again compute and compare the eigenvalues for the perturbed and unperturbed case, plotted in Fig. \ref{fig3}. Notably, the charged case produces negative and diminished eigenvalues for lower overtones, directly stemming from the dominance of electromagnetic effects—making the charged and uncharged scenarios clearly distinguishable in observational data. For the other conditions, namely $N^2 = N^2_{\rm crit} = 96.89$ and $N^2 = 5 N^2_{\rm crit} = 484.44$, the computed eigenvalues are nearly identical, rendering their plots redundant and therefore omitted. \\

\begin{figure}[h!]
\begin{subfigure}{0.5\linewidth}
\centering
\includegraphics[scale = 0.33]{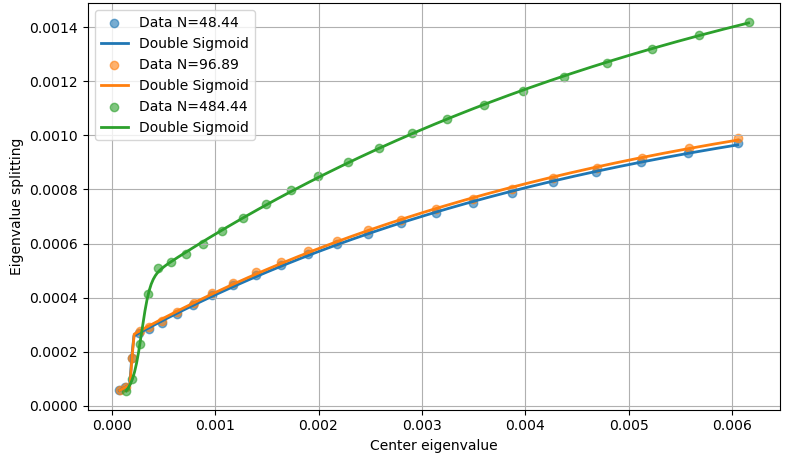}
\subcaption{$\lambda = 2.4 \times 10^{-7}$}
\label{fig4a}
\end{subfigure}
\begin{subfigure}{0.5\linewidth}
\centering
\includegraphics[scale = 0.33]{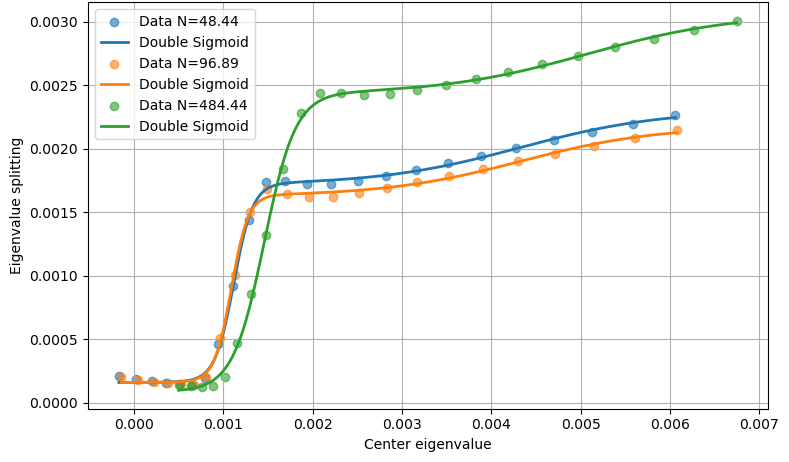}
\subcaption{$\lambda = 1.7 \times 10^{-6}$}
\label{fig4b}
\end{subfigure}
\begin{subfigure}{\linewidth}
\centering
\includegraphics[scale = 0.33]{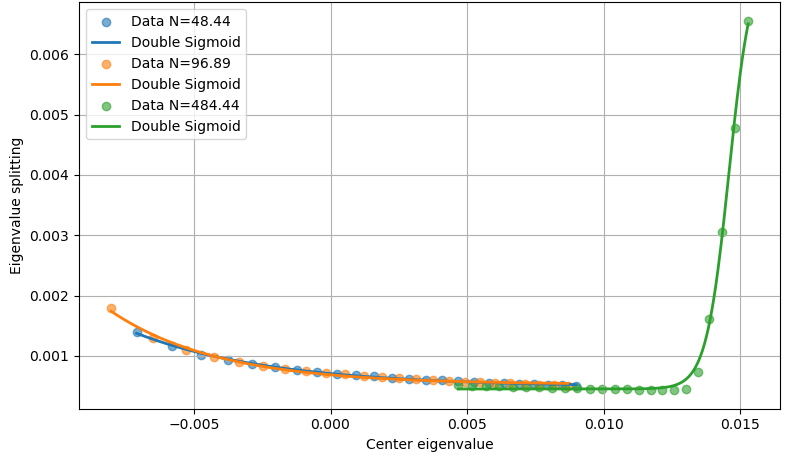}
\subcaption{$\lambda = 2.2 \times 10^{-5}$}
\label{fig4c}
\end{subfigure}
\caption{Eigenvalue splitting vs center eigenvalue for the case of charged cosmic string perturbation for $\lambda \in (10^{-7},10^{-5})$}
\label{fig4}
\end{figure}

The distinction, however, becomes apparent when analysing the eigenvalue center versus splitting plot (Fig. \ref{fig4}), as we clearly see that the data points are different for each condition of $N$. The $N^2 = 5N^2_{\rm crit}$(green) case has the highest eigenvalue splitting, whereas the $N^2 = N^2_{\rm crit}$ (blue) and $N^2 = 0.5N^2_{\rm crit}$ (orange) cases have closely matching values at lower $\lambda$ (Fig. \ref{fig4a}). As $\lambda$ increases, (Fig. \ref{fig4b}), the blue and orange curves initially coincide until a certain eigenvalue center after which the blue curve surpasses the orange in splitting magnitude. The overall trend of the Fig. \ref{fig4} mirrors the uncharged string case, with a peak arising at a lower $\lambda$ (Fig. \ref{fig4a}), moving towards the right (Fig. \ref{fig4b}) eventually forming a minimum (Fig. \ref{fig4c}). The key difference lies in the more rapid evolution of these features relative to the uncharged case, with the peak onset around $\lambda \sim 10^{-6}$. The appearance and movement of these features are critical for distinguishing between different cosmic string properties in GW data. \\

\begin{figure}[h!]
\begin{subfigure}{0.5\linewidth}
\centering
\includegraphics[scale = 0.33]{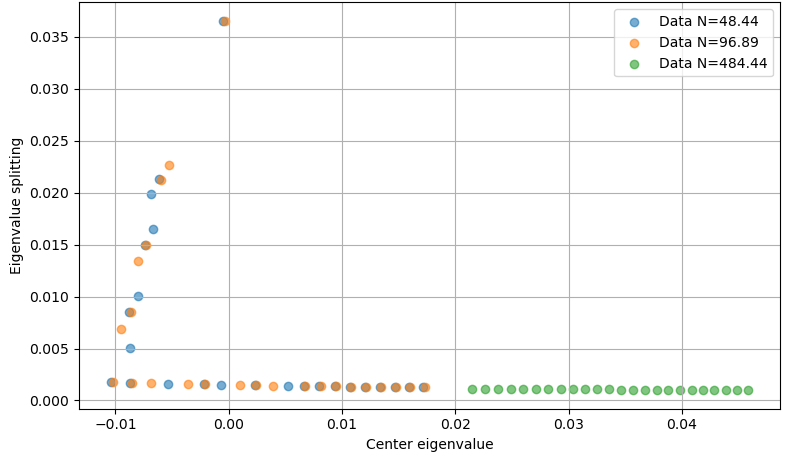}
\subcaption{$\lambda = 8.3 \times 10^{-5}$}
\label{fig5a}
\end{subfigure}
\begin{subfigure}{0.5\linewidth}
\centering
\includegraphics[scale = 0.33]{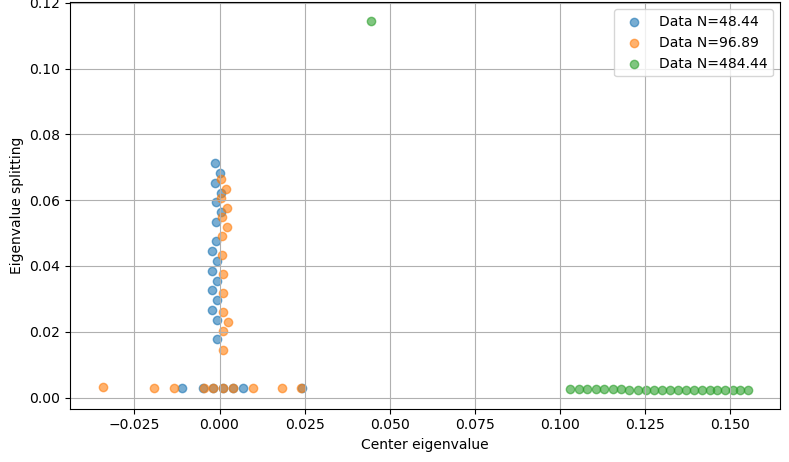}
\subcaption{$\lambda = 4.5 \times 10^{-4}$}
\label{fig5b}
\end{subfigure}
\caption{Eigenvalue splitting vs center eigenvalue for the case of charged cosmic string perturbation for $\lambda \in (10^{-5},10^{-3})$}
\label{fig5}
\end{figure}

\begin{figure}[h!]
\begin{subfigure}{0.5\linewidth}
\centering
\includegraphics[scale = 0.33]{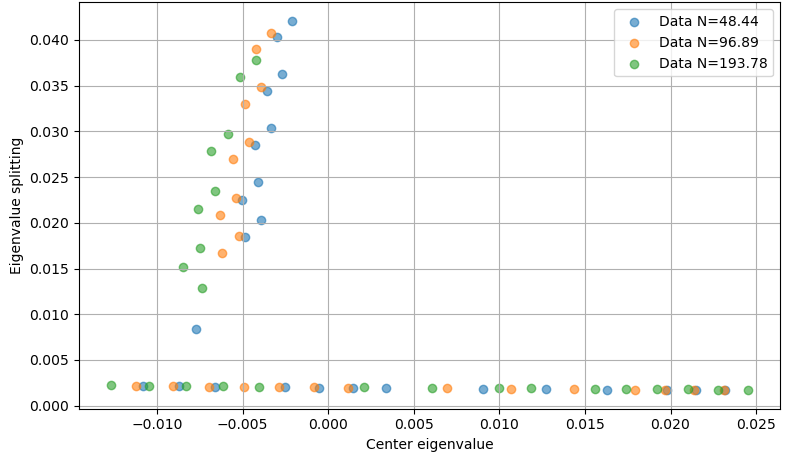}
\subcaption{$\lambda = 10^{-4}$}
\label{fig6a}
\end{subfigure}
\begin{subfigure}{0.5\linewidth}
\centering
\includegraphics[scale = 0.33]{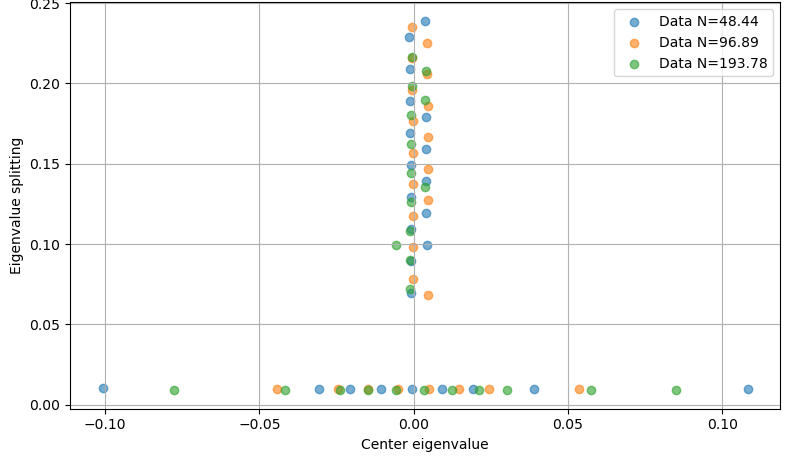}
\subcaption{$\lambda = 1.2 \times 10^{-3}$}
\label{fig6b}
\end{subfigure}
\caption{Eigenvalue splitting vs center eigenvalue for the case of chargedcosmic string perturbation for $\lambda \in (10^{-5},10^{-3})$ and with $N = 193.78$ (green dots)}
\label{fig6}
\end{figure}

As  $\lambda$ approaches $10^{-4}$, the behaviour of the plot changes drastically as visible in Fig. \ref{fig5}. Initially, the data points cluster to form a pronounced peak in the upper left region of Fig. \ref{fig5a}, which subsequently shifts toward zero in terms of the eigenvalue center as shown in Fig. \ref{fig5b}. This vertical clustering around zero represents the phenomenon of avoided crossing or level repulsion \cite{Hund1927, Landau58, Neumann93, Dias22, Lo25} confirming that the perturbation induces this expected effect. It is to be noted that for case where $N^2 = 5N^2_{\rm crit}$, this level repulsion does not manifest, but a lower value of $N$, such as $N^2 = 2N^2_{\rm crit}$, this phenomenon becomes evident, as demonstrated in Fig. \ref{fig6}. This peculiar behaviour highlights the critical influence of $N$ presenting challenges for its precise estimation and constraint.

\section{Results and Discussions}
\label{results}
In this work, we considered a non-rotating Schwarzschild black hole, discussing the Hamiltonian for a zero-momentum photon (\ref{hamiltonian}). We then converted it into operator form (\ref{operator}), (\ref{hamop}) deriving a time independent Schrodinger-like equation (\ref{schrodinger}). Building on this framework, we introduced a one-dimensional perturbation representing the presence of a cosmic string, considering both uncharged (\ref{ucspot}) and charged (\ref{ccspot}) scenarios.\\

Using numerical simulations with fixed parameters --- a black hole of $10M_\odot$, string with energy term $E_s=3M_\odot$, and angular momentum $l = 2$, we computed the eigenvalues across a range of perturbation strengths (Fig. \ref{fig1}). Remarkably, even extremely small perturbations ($ \lambda \sim 10^{-7}$)  induce discernible shifts in the eigenvalue spectrum relative to the unperturbed case, with the smallest detectable effect appearing near $\lambda \lesssim 10^{-10}$. This establishes a critical lower bound for observing cosmic string signatures in QNMs.\\

For the charged cosmic string case, we observe a qualitatively distinct shift where perturbed eigenvalues for lower radial overtones become negative and fall below their unperturbed counterparts (Fig. \ref{fig3}). This contrasts sharply with the uncharged case and arises due to the dominance of the electromagnetic self-interaction component in the effective potential. Another distinctive feature is the constraint on the minimum $\lambda$, with perturbed and unperturbed eigenvalues coinciding for  $\lambda \sim 10^{-7}$  as opposed to $\lambda \sim 10^{-10}$ for uncharged string case. Such behaviour provides a clear observational signature to differentiate between charged and uncharged cosmic strings.\\

Analyzing the QNM eigenvalue centers and splitting reveals unique perturbation dependent patterns. In the uncharged scenario, these follow a characteristic double sigmoid profile (Fig. \ref{fig2}). The charged case similarly exhibits this trend (Fig. \ref{fig4}) but further displays level repulsion phenomena—manifested as avoided crossings in the eigenvalue spectrum—at higher perturbation strengths (Fig. \ref{fig5}, Fig. \ref{fig6}). This phenomenon can serve as a distinctive hallmark in gravitational wave data analysis.\\

Our model's parameters—particularly the perturbation strength $\lambda$, string energy $E_s$, and electromagnetic parameter $N$ for charged strings—exert identifiable influences on the QNM signatures. This model, hence, can be used in an inverse way to estimate the string properties should such QNM signals be detected. For uncharged strings, the inverse problem of extracting $\lambda$ and $E_s$ from observed QNM data is straightforward. However, the presence of the additional $N$ parameter complicates the inversion for charged strings, requiring supplementary electromagnetic constraints for precise parameter estimation.\\

Future work will extend this theoretical framework to other black holes such as rotating (Kerr) and charged (Reissner-Nordström) black holes to enhance model fidelity and observational applicability. As gravitational wave detection precision continues to advance with new instruments, our results underline the potential of QNM analysis not only for probing cosmic strings but also, for identifying a broader class of perturbations influencing black hole ringdown signals.

\providecommand{\newblock}{}

\clearpage

\appendix
\section{Analysis for higher $l$ values}
\label{appendix}

\begin{figure}[h!]
\begin{subfigure}{0.5\linewidth}
\centering
\includegraphics[scale = 0.27]{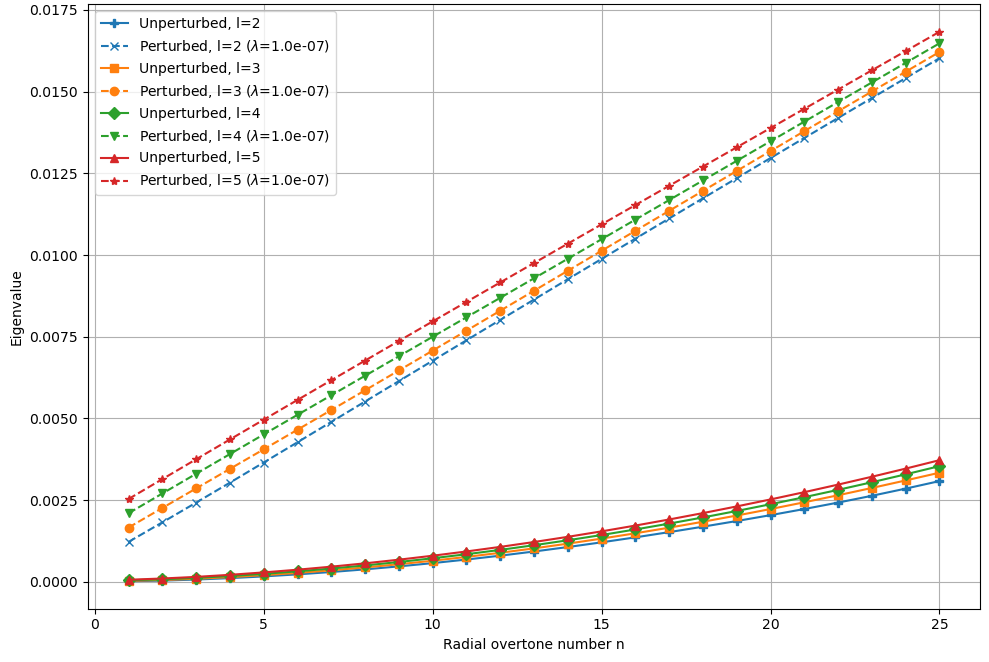}
\subcaption{$\lambda = 10^{-7}$}
\label{fig7a}
\end{subfigure}
\begin{subfigure}{0.5\linewidth}
\centering
\includegraphics[scale = 0.27]{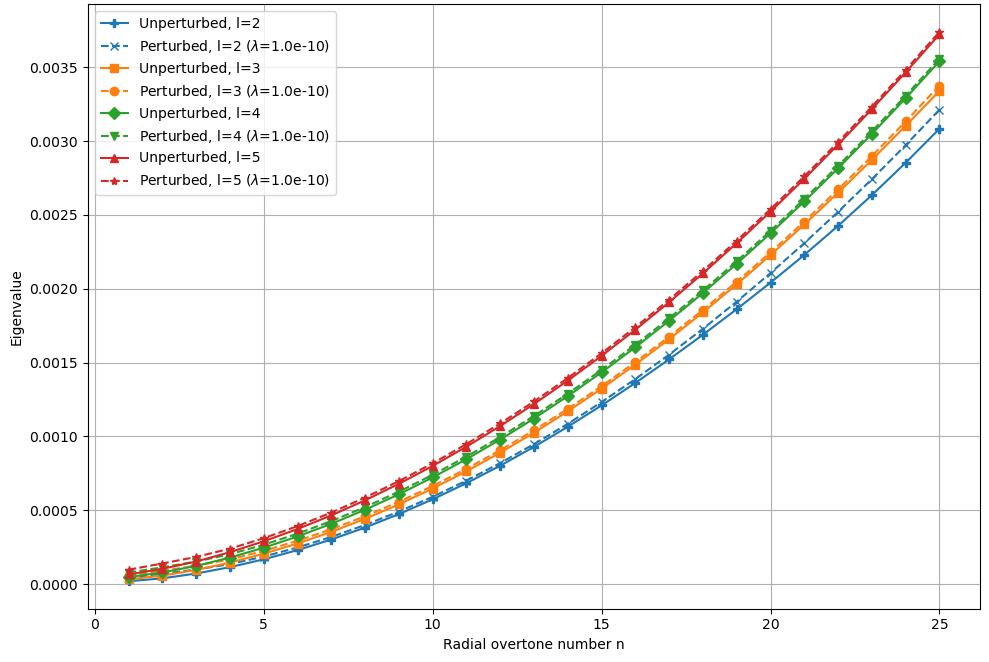}
\subcaption{$\lambda = 10^{-10}$}
\label{fig7b}
\end{subfigure}
\caption{Perturbed vs Unperturbed eigenvalue comparison for uncharged cosmic string perturbation ($l = 2,3,4,5$) }
\label{fig7}
\end{figure}

We now present results for higher angular momentum values $l$ in both uncharged and charged cosmic string scenarios, comparing these with the $l=2$ case analyzed earlier in Section \ref{analysis}. Keeping parameters $M$, $E_s$, and $r$ fixed as before, we simulate the eigenvalues for uncharged string perturbation for $l = 3,4,5$ as shown in Fig. \ref{fig7}. The qualitative features remain consistent, showing a clear eigenvalue shift at $\lambda = 10^{-7}$ (Fig. \ref{fig7a}) and coinciding with the unperturbed eigenvalues at $\lambda = 10^{-10}$ (Fig. \ref{fig7b}). Notably, higher $l$ values correspond to larger eigenvalues, as expected. \\

\begin{figure}[h!]
\begin{subfigure}{0.5\linewidth}
\centering
\includegraphics[width = \textwidth]{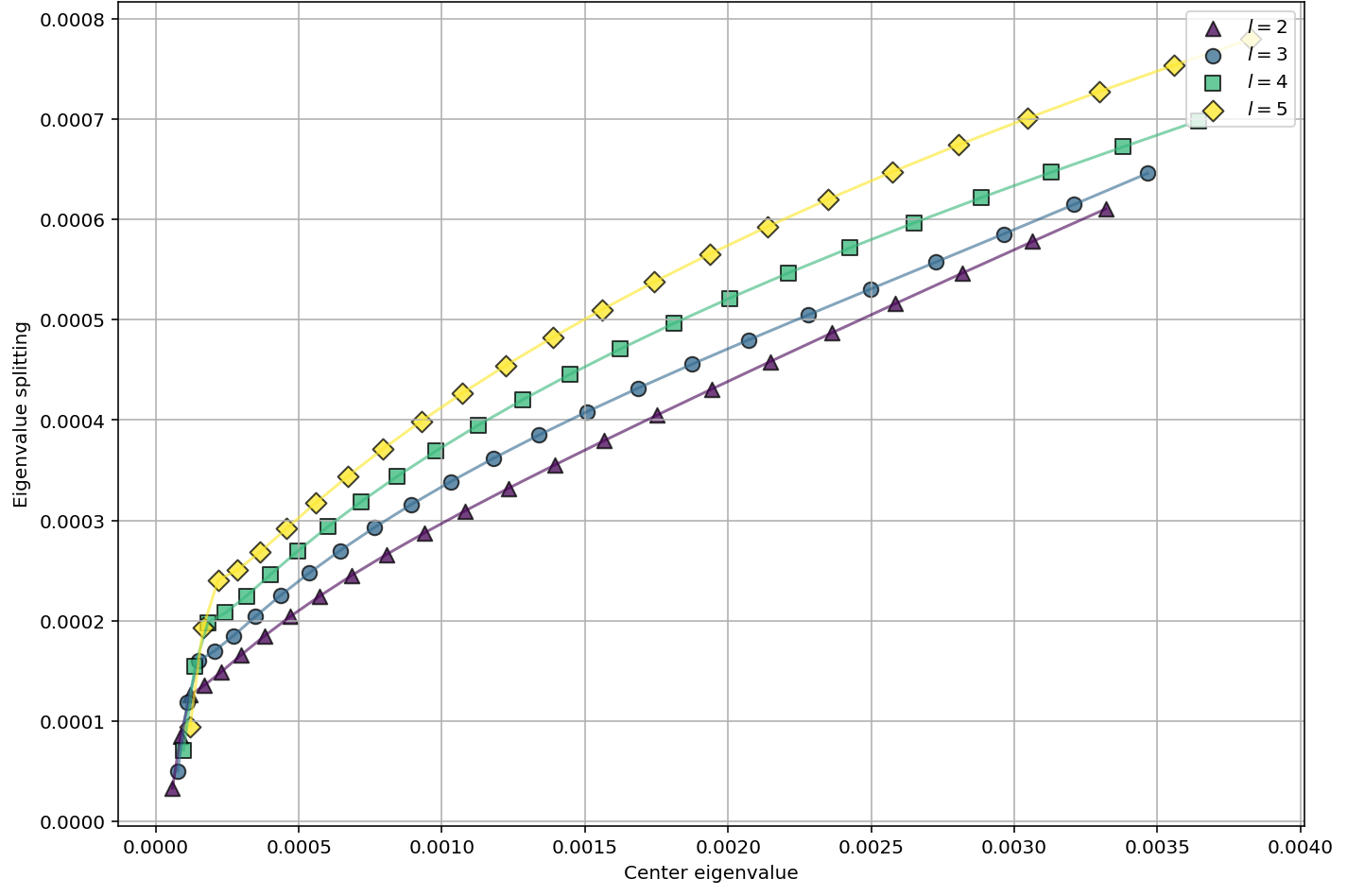}
\subcaption{$\lambda = 10^{-10}$}
\label{fig8a}
\end{subfigure}
\begin{subfigure}{0.5\linewidth}
\centering
\includegraphics[width = \textwidth]{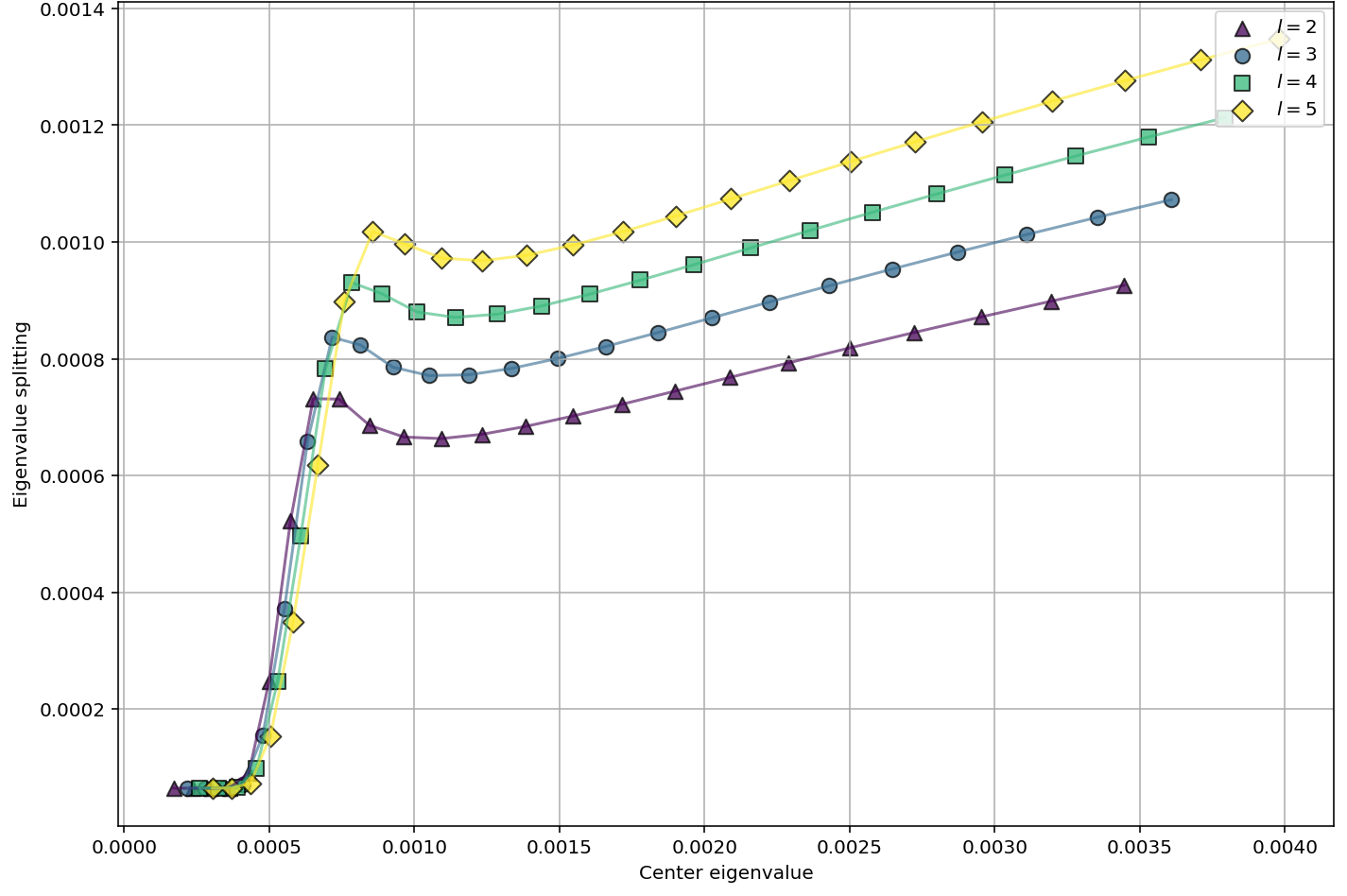}
\subcaption{$\lambda = 10^{-9}$}
\label{fig8b}
\end{subfigure}
\begin{subfigure}{0.5\linewidth}
\centering
\includegraphics[width = \textwidth]{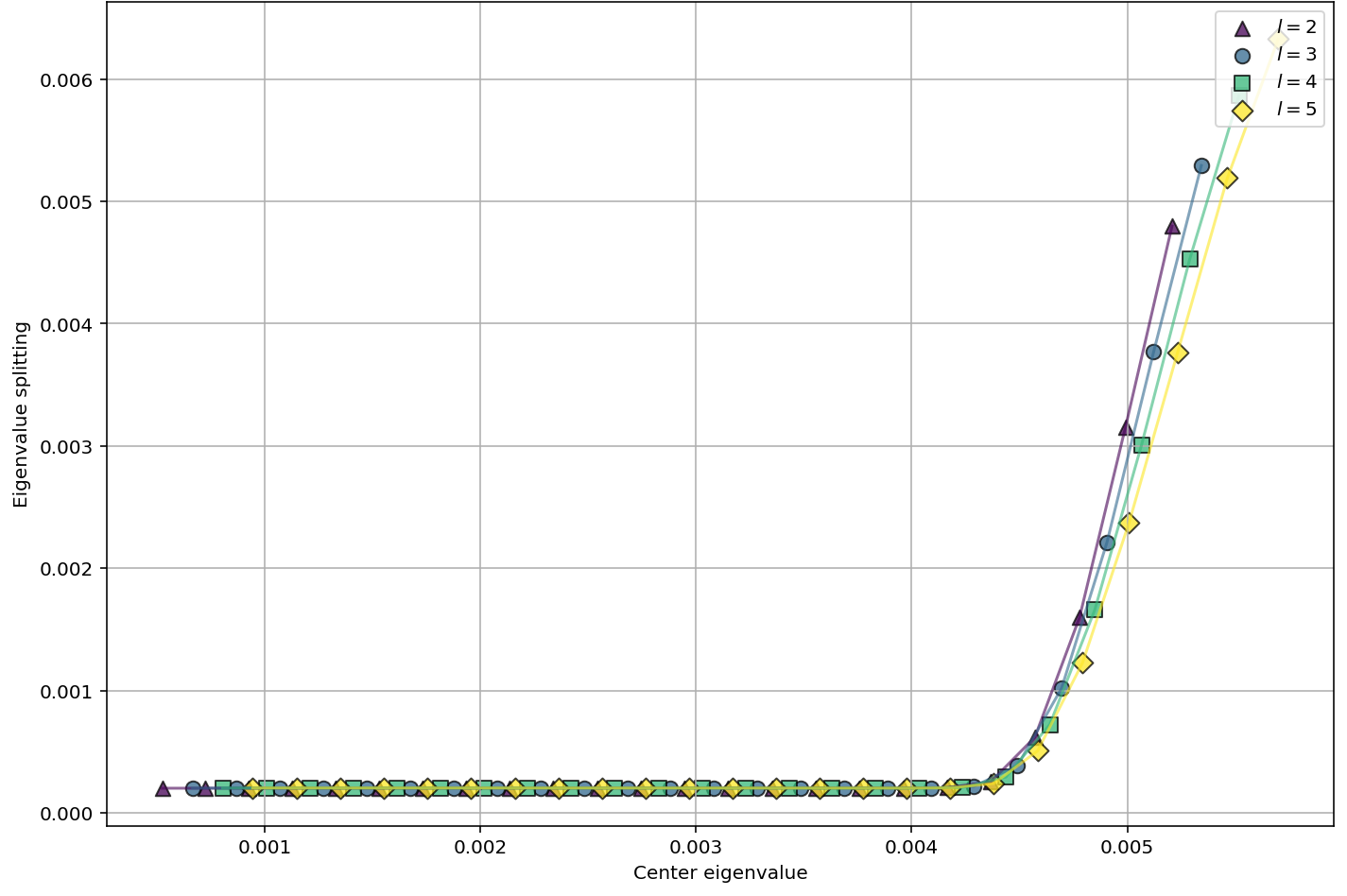}
\subcaption{$\lambda = 10^{-8}$}
\label{fig8c}
\end{subfigure}
\begin{subfigure}{0.5\linewidth}
\centering
\includegraphics[width = \textwidth]{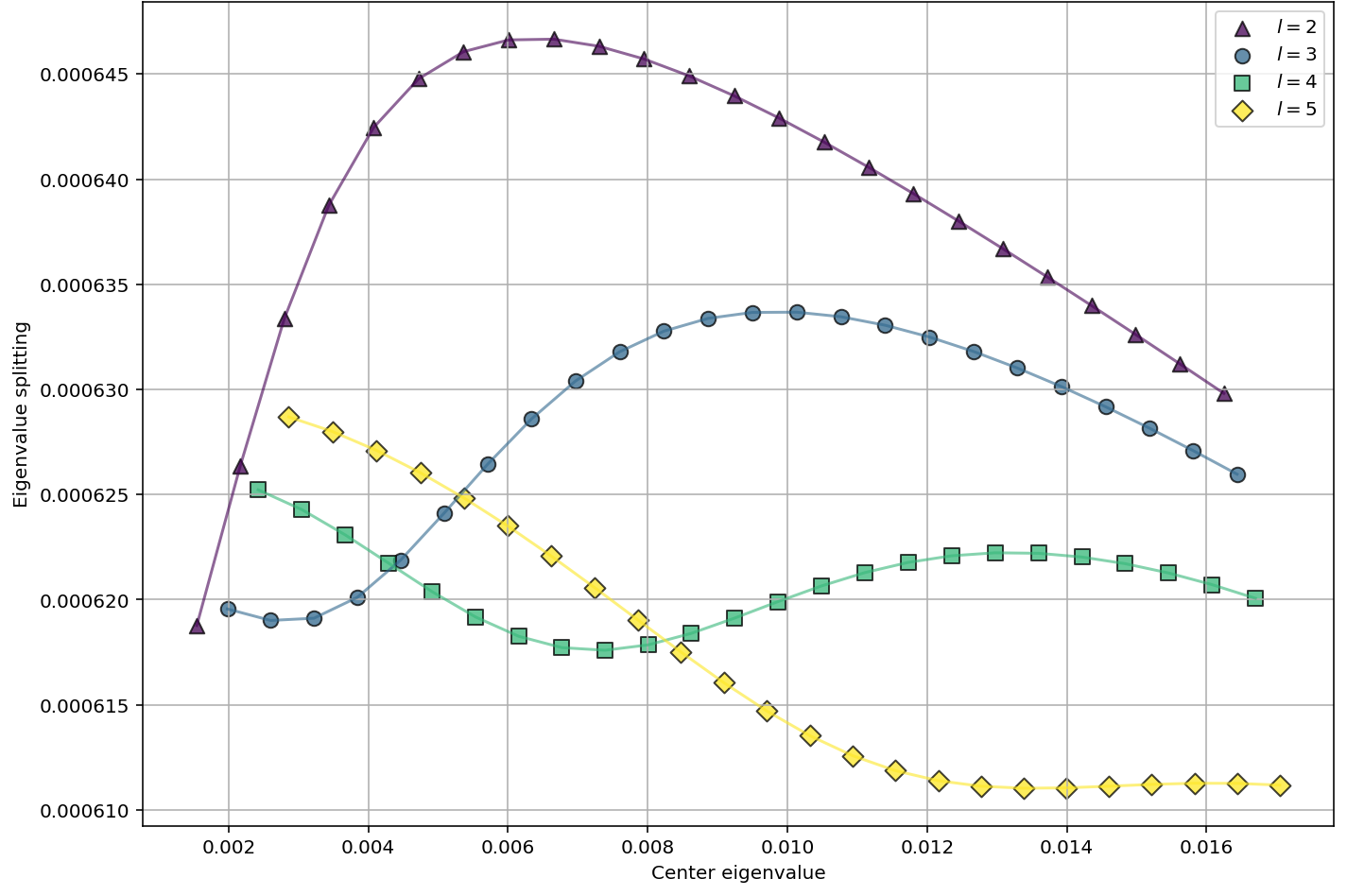}
\subcaption{$\lambda = 10^{-7}$}
\label{fig8d}
\end{subfigure}
\begin{subfigure}{0.5\linewidth}
\centering
\includegraphics[width = \textwidth]{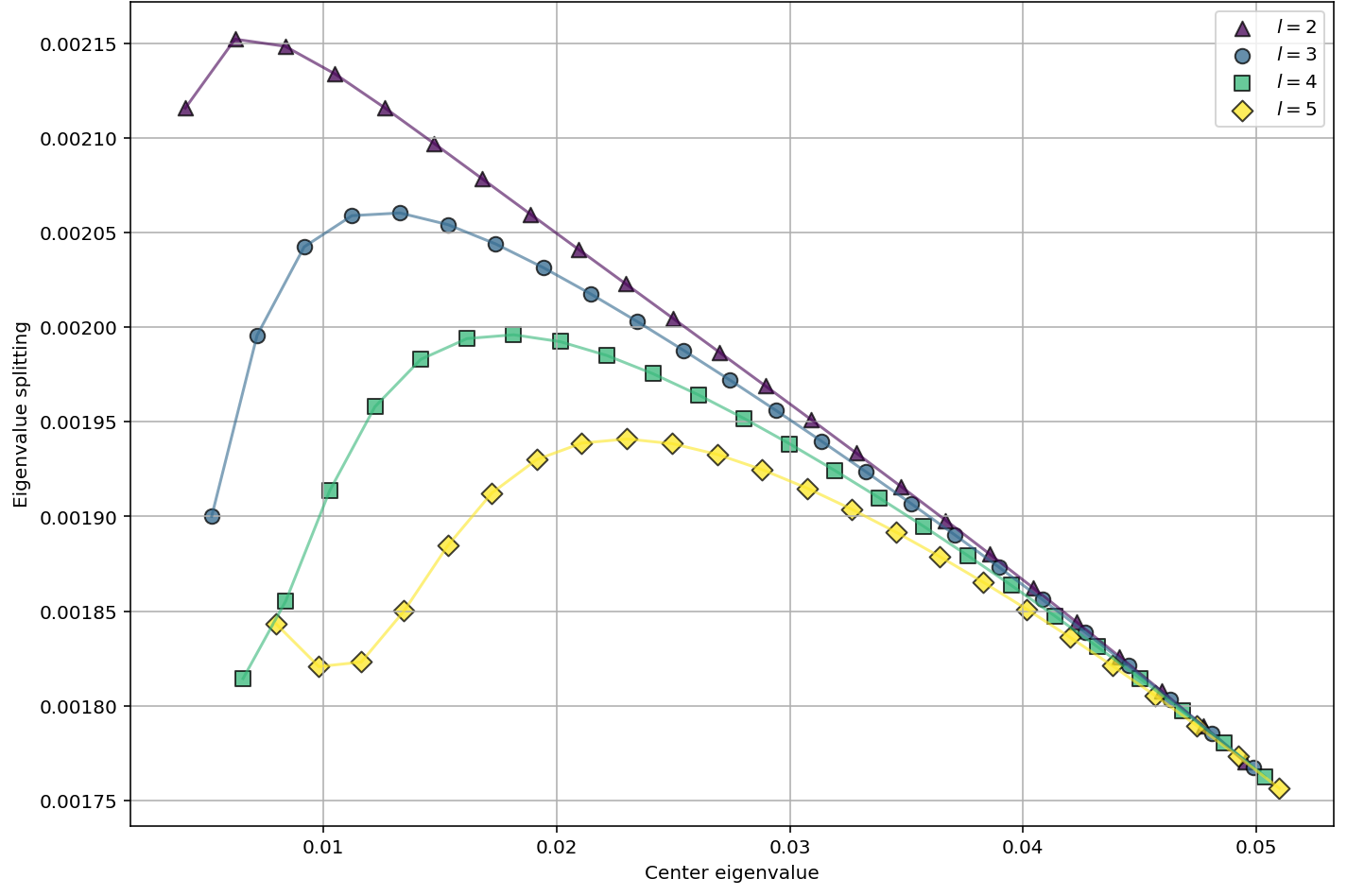}
\subcaption{$\lambda = 10^{-6}$}
\label{fig8e}
\end{subfigure}
\begin{subfigure}{0.5\linewidth}
\centering
\includegraphics[width = \textwidth]{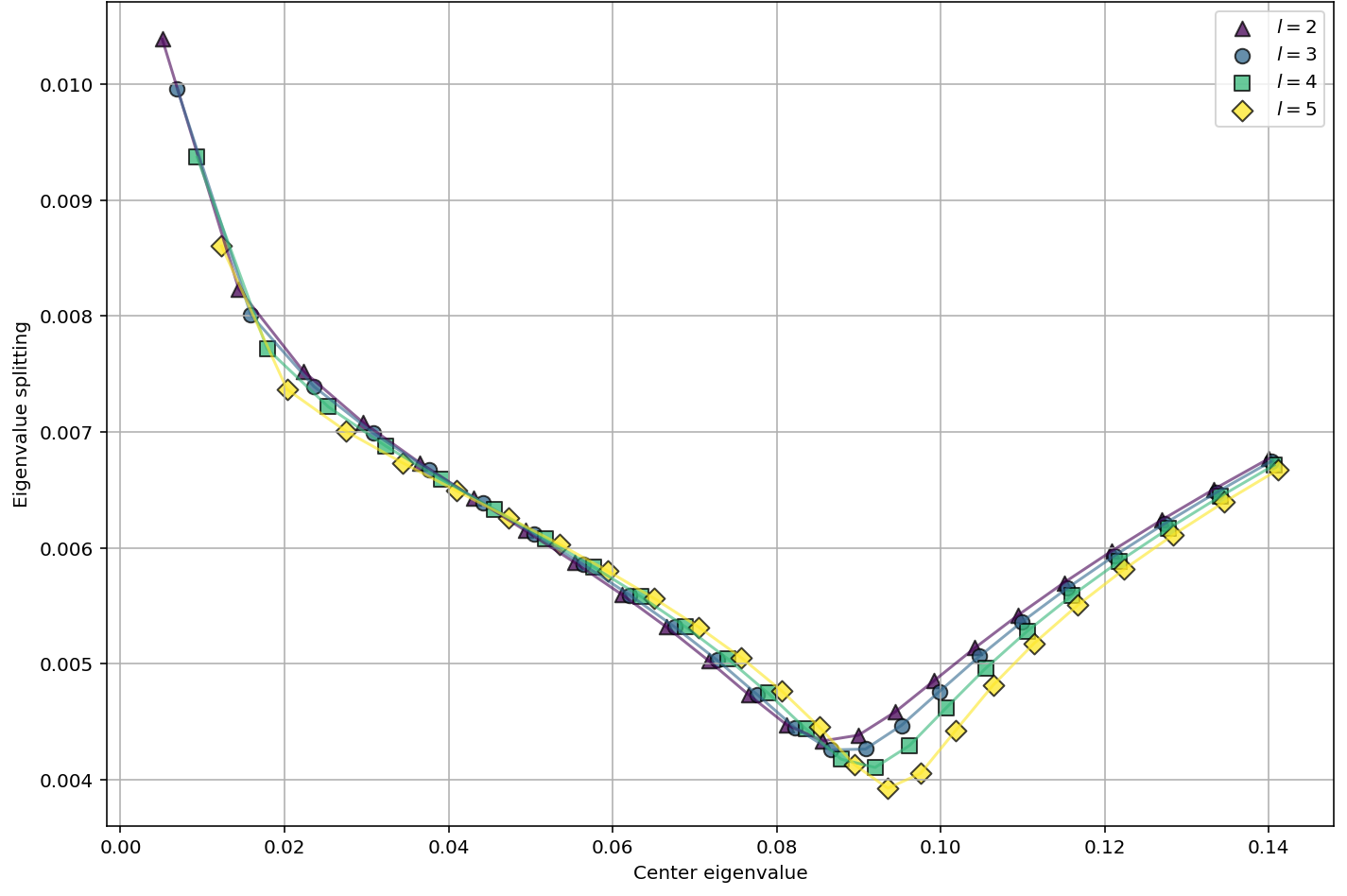}
\subcaption{$\lambda = 10^{-5}$}
\label{fig8f}
\end{subfigure}
\caption{Eigenvalue splitting vs center eigenvalue for the case of uncharged cosmic string perturbation for different lambda values and $l = 2,3,4,5$}
\label{fig8}
\end{figure}

Continuing with the eigenvalue splitting and center calculations for these $l$ values (depicted in Fig. \ref{fig8})  the curve trends are again similar across $l$, with smaller $l$ exhibiting lower splitting in Figs. \ref{fig8a} and \ref{fig8b}. Interestingly, for $\lambda \gtrsim 10^{-8}$this pattern reverses, with smaller $l$ yielding greater splitting (Figs. \ref{fig8c} to \ref{fig8f}). This unusual behaviour lies beyond the present study’s scope and awaits future investigation. \\

\begin{figure}[!htbp]
\begin{subfigure}{0.5\linewidth}
\centering
\includegraphics[scale = 0.27]{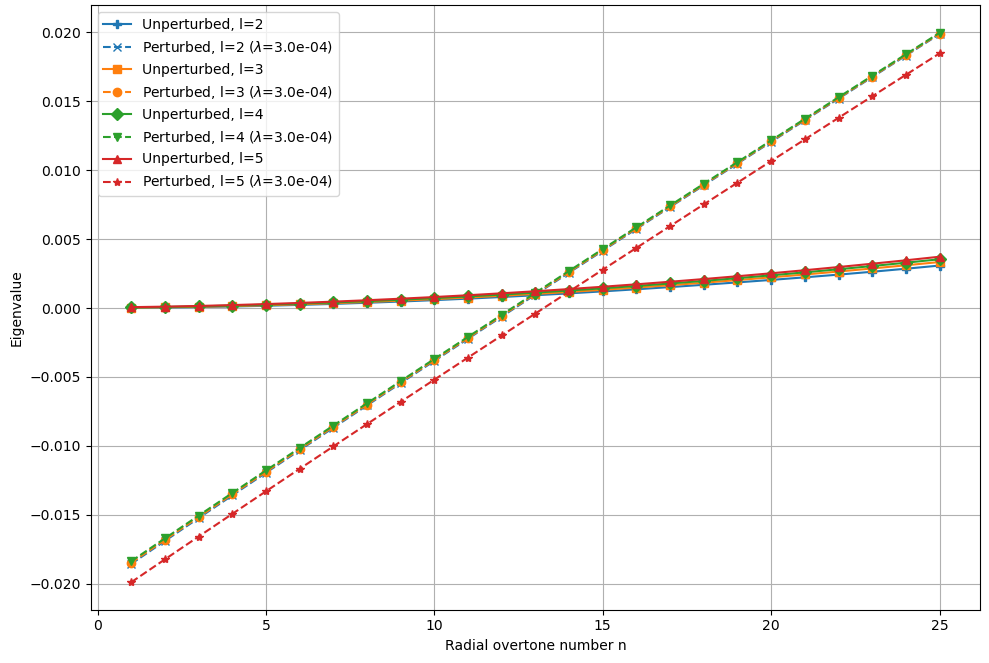}
\subcaption{$\lambda = 10^{-4}$}
\label{fig9a}
\end{subfigure}
\begin{subfigure}{0.5\linewidth}
\centering
\includegraphics[scale = 0.27]{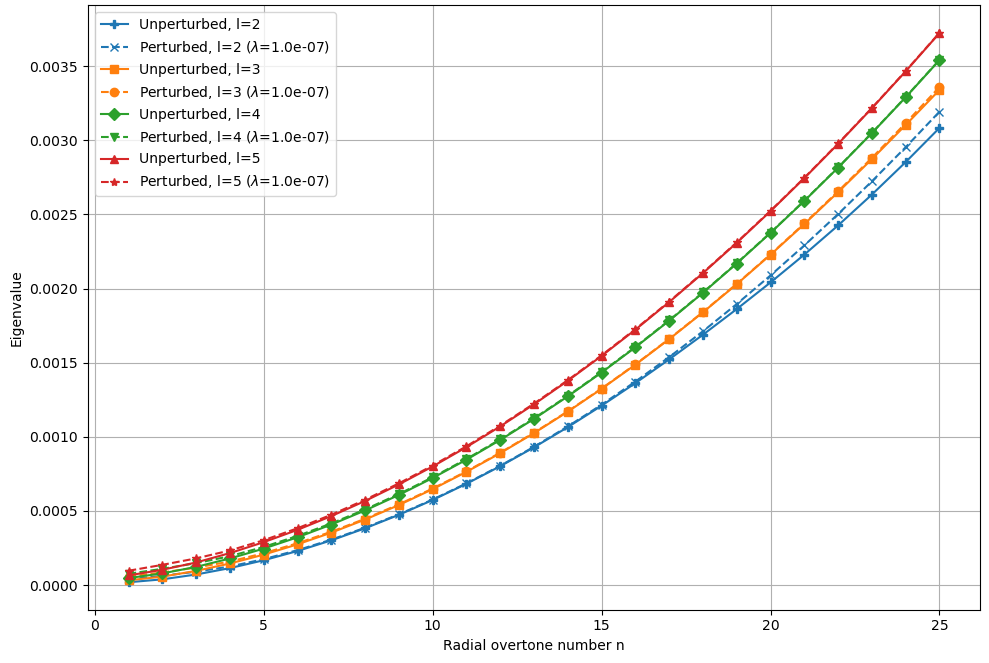}
\subcaption{$\lambda = 10^{-7}$}
\label{fig9b}
\end{subfigure}
\caption{Perturbed vs Unperturbed eigenvalue comparison for charged cosmic string perturbation ($l = 2,3,4,5$) }
\label{fig9}
\end{figure}

In the charged string case, the eigenvalue plots (Fig. \ref{fig9}) show analogous trends at higher $l$, including a visible eigenvalue shift at $\lambda = 10^{-4}$ (Fig. \ref{fig9a}) and convergence with unperturbed spectra at $\lambda = 10^{-7}$ (Fig. \ref{fig9b}). Higher $l$ again, produce slightly higher eigenvalues. \\

\begin{figure}[!htbp]
\begin{subfigure}{0.5\linewidth}
\centering
\includegraphics[scale = 0.27]{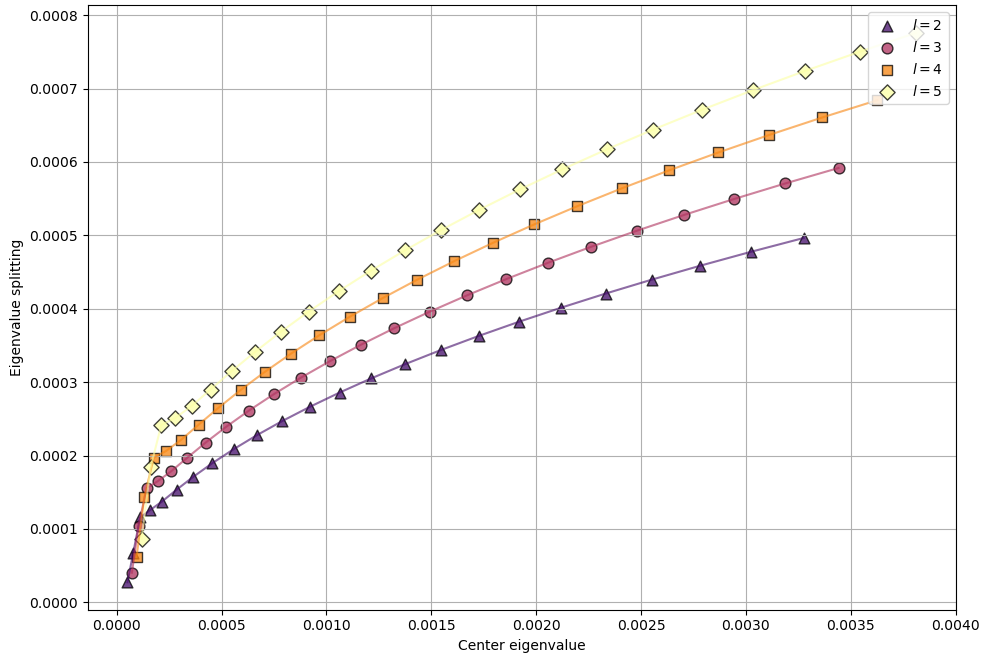}
\subcaption{$\lambda =  10^{-7}$}
\label{fig10a}
\end{subfigure}
\begin{subfigure}{0.5\linewidth}
\centering
\includegraphics[scale = 0.27]{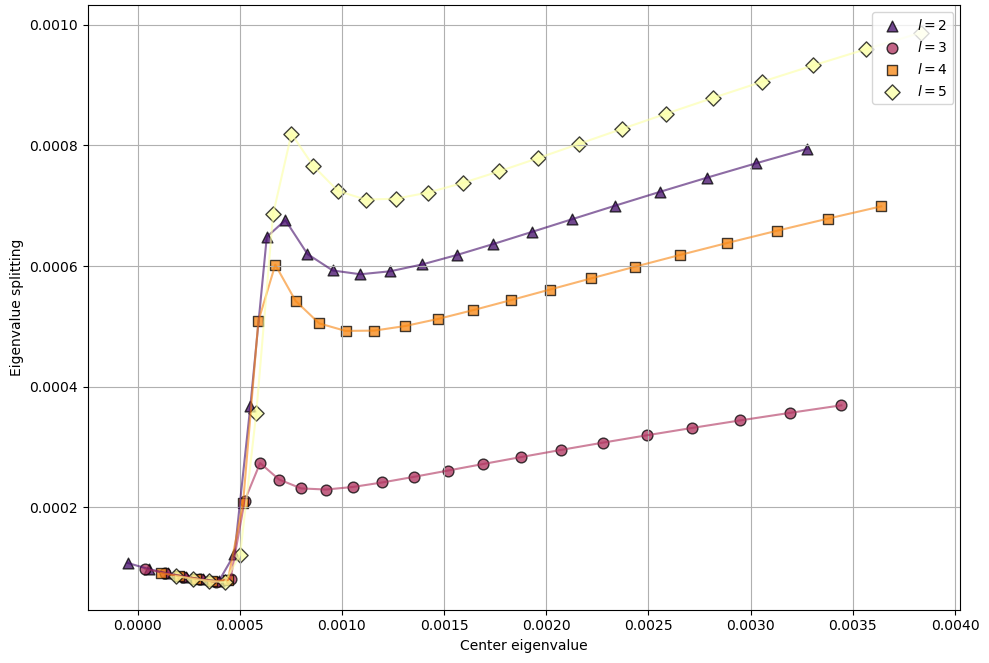}
\subcaption{$\lambda = 10^{-6}$}
\label{fig10b}
\end{subfigure}
\begin{subfigure}{\linewidth}
\centering
\includegraphics[scale = 0.27]{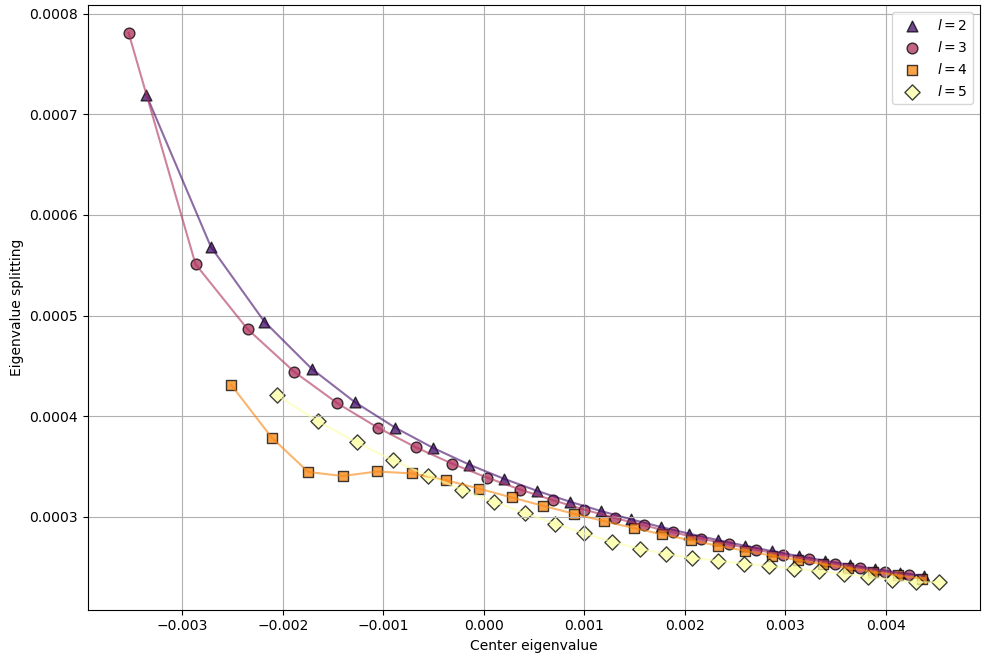}
\subcaption{$\lambda = 10^{-5}$}
\label{fig10c}
\end{subfigure}
\caption{Eigenvalue splitting vs center eigenvalue for the case of charged cosmic string perturbation for $\lambda \in (10^{-7},10^{-5})$ and  $l = 2,3,4,5$}
\label{fig10}
\end{figure}

\begin{figure}[!htbp]
\begin{subfigure}{0.5\linewidth}
\centering
\includegraphics[scale = 0.27]{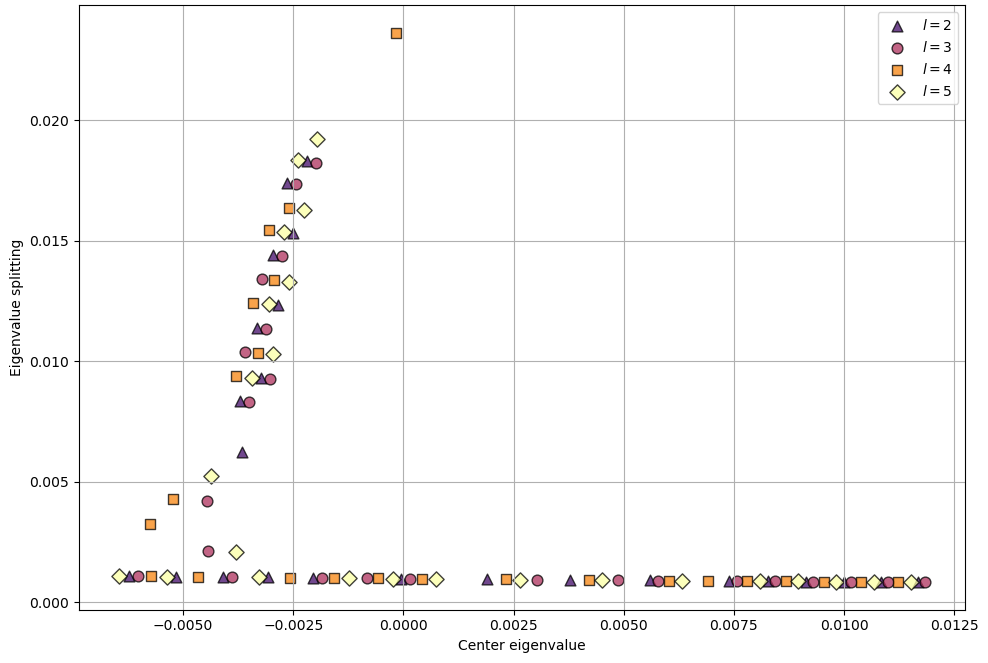}
\subcaption{$\lambda = 8.3 \times 10^{-5}$}
\label{fig11a}
\end{subfigure}
\begin{subfigure}{0.5\linewidth}
\centering
\includegraphics[scale = 0.27]{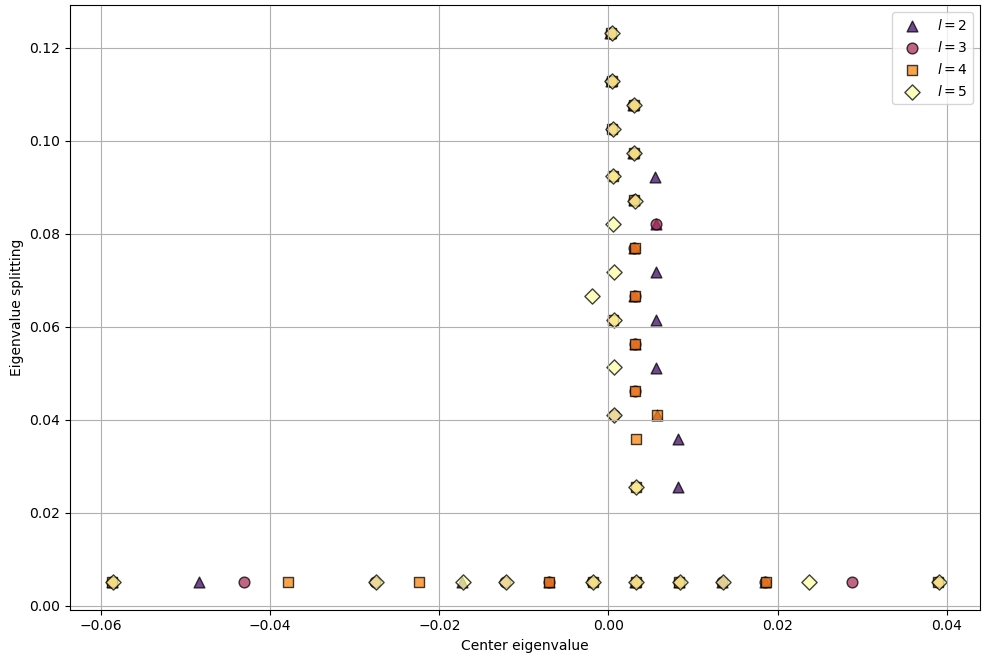}
\subcaption{$\lambda = 4.5 \times 10^{-4}$}
\label{fig11b}
\end{subfigure}
\caption{Eigenvalue splitting vs center eigenvalue for the case of charged cosmic string perturbation for $\lambda \in (10^{-5},10^{-3})$ and $l = 2,3,4,5$}
\label{fig11}
\end{figure}

Furthermore,  Fig. \ref{fig10} displays splitting versus centers for $l = 2,3,4,5$ in the charged scenario, mirroring the uncharged case results. Consistent with previous results, we notice that for $\lambda \gtrsim 10^{-4}$,  lower $l$ produces larger splitting, demonstrating that this behaviour is not unique to uncharged strings.  Level repulsion phenomena also occur for higher $l$ at the same $\lambda$, as illustrated in Fig. \ref{fig11}.\\

We point out that these results correspond to the condition $N^2 = 0.5 N^2_{crit} = 48.44$, and the analysis for the other two conditions, $N^2 = N^2_{\rm crit} = 96.89$ and $N^2 = 5 N^2_{\rm crit} = 484.44$ have been omitted as they show similar trends, albeit with differing eigenvalues, splitting and centers. To summarize, changes in $l$ affect eigenvalues, splitting and centers quantitatively, yet the fundamental behaviour of the plots remains consistent for both uncharged and charged string perturbations.

\end{document}